\documentclass[preprint,12pt]{elsarticle}

\usepackage{amssymb}
\usepackage{mathtools}
\usepackage{amsfonts}
\usepackage{todonotes}

\journal{}

\newtheorem{thm}{Theorem}
\newtheorem{lem}[thm]{Lemma}
\newtheorem{prop}[thm]{Proposition}

\newdefinition{rmk}{Remark}
\newproof{prf}{Proof}
\newproof{pot}{Proof of Theorem \ref{thm2}}

\newcommand{\N}{\mathbb{N}}
\newcommand{\Q}{\mathbb{Q}}
\newcommand{\R}{\mathbb{R}}

\newlength{\problemoffset}
\newcommand{\decision}[3]{
\begin{list}{}{
\setlength{\leftmargin}{\problemoffset}
\setlength{\rightmargin}{\problemoffset}
\setlength{\parsep}{0pt}
\setlength{\itemsep}{2pt}
\setlength{\topsep}{\itemsep}
\setlength{\partopsep}{\itemsep}
}
\item
{\textsc{#1}}
\item
{{\bf INSTANCE:} #2}
\item
{{\bf QUESTION:} #3}
\end{list}
}

\begin{document}

\begin{frontmatter}



\title{The Complexity of the $K$th Largest Subset Problem and Related
  Problems}


\author[cachan]{Christoph Haase\fnref{fn1}}
\address[cachan]{LSV, CNRS \&
  ENS Cachan, Universit\'e Paris-Saclay, France}
\fntext[fn1]{Supported by Labex Digicosme, Univ. Paris-Saclay, project
  VERICONISS.}

\author[oxford]{Stefan Kiefer\fnref{fn2}}
\address[oxford]{Department of Computer Science, University of Oxford, UK}
\fntext[fn2]{Supported by a University Research Fellowship of the Royal Society.}

\begin{abstract}
We show that the \textsc{$K$th largest subset} problem and the
\textsc{$K$th largest $m$-tuple} problem are in PP and hard for PP
under polynomial-time Turing reductions. Several problems from the
literature were previously shown NP-hard via reductions from those two
problems, and by our main result they become PP-hard as well. We also
provide complementary PP-upper bounds for some of them.
\end{abstract}


\end{frontmatter}

\section{Introduction} \label{sec-intro}

The following two problems are listed in Garey and Johnson's classical
compendium on the theory of NP-completeness~\cite[p.\ 225]{GJ79} as
problems [SP20] and [SP21], respectively:

\decision{$K$th largest subset problem}
{A set $X=\{x_1,\ldots, x_m\}\subseteq \N$ and $K,B\in \N$.}
{Is $\#\left\{ Y\subseteq X : \sum_{x\in Y} x \le B \right\} \ge K$ ?}

\vspace{2mm}

\decision{$K$th largest $m$-tuple problem}
{Finite sets $X_1,\ldots, X_m\subseteq \N$ and $K,B\in \N$.}
{Is $\#\left\{ (x_1,\ldots,x_m)\in X_1\times \cdots X_m : \sum_{i=1}^m x_i
  \ge B \right\} \ge K$~?}
The first problem was initially introduced by Johnson and
Kashdan~\cite{JK78} and the second by Johnson and
Mizoguchi~\cite{JM78}.  Garey and Johnson~\cite{GJ79} state that both
problems are NP-hard under polynomial-time Turing reductions but
mention that their membership in NP, and, thus, their NP-completeness,
is open. For the NP-hardness proof of the \textsc{$K$th largest
  subset} problem, see Theorem~6 in the paper of Johnson and
Kashdan~\cite{JK78}, or the discussion provided by Garey and
Johnson~\cite[p.\ 115]{GJ79}. It is unknown whether this problem is
also NP-hard under polynomial-time many-one
reductions~\cite[p.\ 148]{HS11}.
The \textsc{$K$th largest subset} problem has its name for historical reasons:
note that the the problem is about the $K$th \emph{smallest} subset.

Both problems have repeatedly been used as starting points for proving
other problems NP-hard. In this note, we show that the \textsc{$K$th
  largest subset} and the \textsc{$K$th largest $m$-tuple} problem are
complete for the complexity class PP. More precisely, they are in PP,
and they are hard for PP under polynomial-time Turing reductions. The
complexity class PP~\cite{Gill77:PP} can be defined as the class of
languages~${L\subseteq \Sigma^*}$ that have an NP Turing machine~$M_L$
such that for all words~${w\in \Sigma^*}$ one has ${w \in L}$ if and
only if more than half of the computation paths of~$M_L$ on~$w$ are
accepting. The class~PP includes~NP~\cite{Gill77:PP}.  Closely related
is the function class~\#P, which consists of those
functions~$f:\Sigma^* \to \mathbb{N}$ for which there exists an NP
Turing machine~$M_f$ such that for all words~$w\in \Sigma^*$ the
function value $f(w)$ is equal to the number of accepting computation
paths of~$M_f$ on~$w$.  The class~PP is remarkably powerful: Toda's
Theorem~\cite{Toda91} states that P$^\text{PP}$ (which equals
P$^\text{\#P}$) contains the polynomial-time hierarchy PH.  This
means, in particular, that if the \textsc{$K$th largest subset} (or
\textsc{$m$-tuple}, resp.) problem were in NP then the polynomial-time
hierarchy would collapse to P$^\text{NP}$. This can be taken as
evidence (under standard assumptions from complexity theory) that the
problems considered in this note are not in NP and, hence, are harder
than NP.

By an application of the main result of this note, it follows that
several other problems for which reductions from the \textsc{$K$th
  largest subset} (or \textsc{$m$-tuple}) problem have been
constructed are not only NP-hard under polynomial-time Turing
reductions, but in fact PP-hard under polynomial-time Turing
reductions. We give examples of such problems in
Section~\ref{sec-other-problems}.


\section{Main Result} \label{sec-pp-completeness}

The class~PP can equivalently be characterised as the class of
languages~$L \subseteq \Sigma^*$ that have a polynomial-time bounded
deterministic Turing machine~$M$ with access to a fair coin such that
for all $w \in \Sigma^*$ we have $w \in L$ if and only if $M$
accepts~$w$ with probability at least~$1/2$.  This follows from
Theorem~4.4 of Simon~\cite{Simon75}, and in particular from the fact
that PP is closed under complement~\cite{Simon75}, see also the work
of Gill~\cite{Gill77:PP}. However, in some proofs it will be more
convenient if the Turing machine has access to \emph{biased} rather
than fair coins, and if the biases (rational numbers between 0~and~1)
can depend on the input.  More precisely, a \emph{probabilistic
  polynomial-time bounded Turing machine (PP-TM)} is a polynomial-time
bounded Turing machine that takes an input~$w$ and then operates in
two phases:
\begin{enumerate}
\item Depending only on~$w$, it deterministically computes finitely
  many rational numbers $b_1, \ldots, b_k$ with $b_i \in (0,1)$ for
  all $i \in \{1, \ldots, k\}$.  Each~$b_i$ is represented as a
  quotient of integers written in binary.  We view the~$b_i$ values as
  \emph{biases} of biased coins.
\item It acts like a nondeterministic polynomial-time bounded Turing
  machine, but the nondeterminism is resolved probabilistically using
  biased coins with biases $b_1, \ldots, b_k$ that were computed in
  the first phase.  More concretely, besides deterministic
  instructions, the Turing machine has instructions of the form
  \[
  \texttt{if coin}(b_i) \texttt{ then goto c1 else goto c2}
  \]
  where $b_i$ is a bias computed in the first phase.  The semantics of
  this operation is the natural one: with probability~$b_i$ the Turing
  machine continues at~\texttt{c1}, otherwise (i.e., with
  probability~$1-b_i$) it continues at~\texttt{c2}.
\end{enumerate}
We have the following lemma.
\begin{lem} \label{lem-PP-threshold}
Let $L \subseteq \Sigma^*$ and $M$ be a PP-TM. Furthermore, let $\tau:
\Sigma^* \to (0,1)$ be a polynomial-time computable function returning
a rational number encoded as a quotient of integers in binary such
that for all $w \in \Sigma^*$ we have $w \in L$ if and only if $M$
accepts~$w$ with probability at least~$\tau(w)$. Then $L \in
\text{PP}$.
\end{lem}
\begin{prf}
The proof consists of three steps:
\begin{enumerate}
\item We show that we can, without loss of generality, take $\tau(w) =
  1/2$ for all $w \in \Sigma^*$.
\item We construct from~$M$ an NP Turing machine~$M'$ such that for
  all ${w\in \Sigma^*}$ we have ${w \in L}$ if and only if at least
  half of the computation paths of~$M'$ on~$w$ are accepting.
\item We argue that the existence of~$M'$ implies membership of~$L$
  in~PP.
\end{enumerate}
In the first step of the proof, we show that we can, without loss of
generality, take $\tau(w) = 1/2$ for all $w \in \Sigma^*$.  Indeed,
given $M$~and~$\tau$, construct a PP-TM~$M'$ as follows. Given $w \in
\Sigma^*$, the PP-TM~$M'$ first computes $\tau := \tau(w)$.  If $\tau
< 1/2$, it computes $p := (1/2 - \tau) / (1 - \tau)$, and accepts
immediately with probability~$p$.  Otherwise (i.e., with
probability~$1-p$) the PP-TM~$M'$ simulates~$M$.  It follows that
\begin{align*}
  \Pr(M' \text{ accepts } w) = p + (1-p) \cdot \Pr(M \text{ accepts }
w).
\end{align*}
A simple calculation shows that $\Pr(M \text{ accepts } w) \ge
\tau(w)$ if and only if $\Pr(M' \text{ accepts } w) \ge 1/2$.

Similarly, if $\tau > 1/2$, the PP-TM~$M'$ computes $p := 1/(2\tau)$,
and rejects immediately with probability~$1-p$.  Otherwise (i.e., with
probability~$p$) the PP-TM~$M'$ simulates~$M$.  It follows that
\begin{align*}
  \Pr(M' \text{ accepts } w) = p \cdot \Pr(M \text{ accepts } w).
\end{align*}
Again, we have $\Pr(M \text{ accepts } w) \ge \tau(w)$ if and only if
$\Pr(M' \text{ accepts } w) \ge 1/2$.

Consequently, for the remainder of the proof we can assume that for
all $w \in \Sigma^*$ we have $w \in L$ if and only if $M$ accepts~$w$
with probability at least~$1/2$.  Furthermore, we can ensure that for
any $w \in \Sigma^*$ all computation paths of~$M$ on~$w$ involve the
same number, say $p_w$, of accesses to a coin.

In the second step of the proof, we construct from~$M$ an NP Turing
machine~$M'$ such that for all ${w\in \Sigma^*}$ we have ${w \in L}$
if and only if at least half of the computation paths of~$M'$ on~$w$
are accepting.
Given a word $w \in \Sigma^*$, the Turing machine~$M'$ computes, like
the PP-TM~$M$, the set $B$ of biases.  Then $M'$ computes a multiple,
say~$q_w$, of the denominators of the biases in~$B$. Subsequently,
$M'$ simulates~$M$, except that when it hits an instruction
\[
 \texttt{if coin}(b) \texttt{ then goto c1 else goto c2}
\]
then $M'$ takes $b \cdot q_w$ nondeterministic branches to \texttt{c1}
and $(1-b) \cdot q_w$ nondeterministic branches to \texttt{c2}.  By
the definition of~$q_w$ the numbers $b \cdot q_w$ and $(1-b) \cdot
q_w$ are natural numbers.  They might be of exponential magnitude, but
their representation in binary is of polynomial length.  Using this
representation, the Turing machine~$M'$ can implement the $b \cdot
q_w$ and $(1-b) \cdot q_w$ nondeterministic branches using cascades of
binary nondeterministic branches.  By this construction, we have for
any $w \in \Sigma^*$ that the total number of computation paths
of~$M'$ on~$w$ equals $(q_w)^{p_w}$ and, furthermore, if $M$
accepts~$w$ with probability~$x$ then $x \cdot (q_w)^{p_w}$
computation paths of~$M'$ on~$w$ are accepting.  We conclude that we
have $w \in L$ if and only if at least half of the computation paths
of~$M'$ on~$w$ are accepting.

For the third and final step of the proof, 
construct an NP Turing
machine~$M''$ from~$M'$ by swapping accepting and rejecting states.
Then we have $w \not\in L$ if and only if \emph{more than half} of the
computation paths of~$M''$ on~$w$ are accepting.  Hence, using the definition of~PP, the complement
of~$L$ is in~PP.  But PP is closed under complement,
see~\cite{Gill77:PP}, thus $L \in \text{PP}$. \qed
\end{prf}

In order to show PP-hardness of the \textsc{$K$th largest subset} and
the \textsc{$K$th largest $m$-tuple} problem, we employ
\#P-completeness of a variant of the classical \textsc{SubsetSum}
problem. Given a set $X=\{ x_1,\ldots,x_m \}\subseteq \mathbb{N}$ and
$T\in \mathbb{N}$, \textsc{\#SubsetSum} asks for determining the
number of subsets $Y$ of $X$ whose elements sum up to $T$, i.e.,
computing
\[
\#\left\{ Y \subseteq X : \sum_{x\in Y} x = T \right\}.
\]
Building upon the results of Hunt et al.~\cite{HMRS98}, Faliszewski
and Hemaspaandra \cite[p.\ 104]{Hemaspaandra09} derive that
\textsc{\#SubsetSum} is \#P-complete, where \#P-hardness holds under
parsimonious many-one reductions. As defined by Simon~\cite{Simon75},
given functions $f,g: \Sigma^* \to \mathbb{N}$ we say that $f$
\emph{parsimoniously reduces to} $g$ if there is a polynomial-time
computable function $h: \Sigma^*\to \Sigma^*$ such that $f(w)=g(h(w))$
for all $w\in \Sigma^*$. For subsequent reference, let us capture the
result of Faliszewski and Hemaspaandra09~\cite{Hemaspaandra09} by the
following proposition.
\begin{prop}
  \label{prop:sharp-subsetsum}
  \textsc{\#SubsetSum} is \textsc{\#P}-complete, and in particular
  \textsc{\#P}-hard under parsimonious many-one reductions.
\end{prop}

We are now prepared to prove our main result. In the remainder of this
note, whenever we refer to PP-completeness, PP-hardness holds under
polynomial-time Turing reductions.  In the statements of the theorems
we make this explicit to avoid a confusion of casual readers.

\begin{thm} \label{thm-main}
  The \textsc{$K$th largest subset} problem and the \textsc{$K$th
    largest $m$-tuple} problem are PP-complete under polynomial-time
  Turing reductions.
\end{thm}
\begin{prf}
  First, we prove membership in PP for both problems. Given an
  instance $X=\{x_1,\ldots, x_m\}\subseteq \mathbb{N}$ and $B, K\in
  \mathbb{N}$ of the \textsc{$K$th largest subset} problem, we can
  construct a PP-TM $M$ that first chooses every subset ${Y\subseteq
    X}$ uniformly at random by selecting every~$x_i$ with
  probability~$1/2$. Subsequently, $M$ accepts if and only if $\sum_{x
    \in Y} x\le B$. It follows that the probability that $M$ accepts
  is equal to
  \begin{align*}
    \#\left\{ Y \subseteq X : \sum_{x\in Y} x \le B\right \} \big/ 2^m.
  \end{align*}
  Hence, $\#\{ Y\subseteq X : \sum_{x\in Y}x \le B \} \ge K$ if and
  only if $M$ accepts with probability at least $K/2^m$. By
  Lemma~\ref{lem-PP-threshold}, the \textsc{$K$th largest subset} problem is in~PP.

  For the \textsc{$K$th largest $m$-tuple} problem, the construction
  is similar: a PP-TM~$M$ selects an $m$-tuple $(x_1,\ldots,x_m)\in
  X_1\times \cdots \times X_m$ uniformly at random, i.e., each
  $m$-tuple with probability $1 \big/ \prod_{i=1}^m (\# X_i)$. By the
  characterisation of PP in Lemma~\ref{lem-PP-threshold}, this can be
  achieved by independently choosing some $x_i\in X_i$ with
  probability $1/\# X_i$ for every $1\le i\le m$.  Then, $M$ accepts
  if and only if $\sum_{i=1}^m x_i \ge B$.  Hence, $\#\left\{
  (x_1,\ldots,x_m)\in X_1\times \cdots \times X_m : \sum_{i=1}^m x_i
  \ge B \right\} \ge K$ if and only if $M$ accepts with probability at
  least $K \big/ \prod_{i=1}^m (\# X_i)$. 
  An application of Lemma~\ref{lem-PP-threshold} shows that
  the \textsc{$K$th largest $m$-tuple} problem is in~PP as well.

  In order to show hardness for PP, we give polynomial-time Turing
  reductions from the canonical PP-complete problem
  \textsc{MajSAT}~\cite{Gill77:PP,Pap-book}. This problem asks, given
  a Boolean formula~$\psi$ over $n$ Boolean variables $b_1,\ldots,
  b_n$, does a majority (i.e., at least $2^{n-1}+1$) of the variable
  assignments satisfy~$\psi$. Since \#SAT, i.e., the task of computing
  the number of satisfying assignments of a Boolean formula, is
  \#P-complete~\cite{Pap-book}, it follows from
  Proposition~\ref{prop:sharp-subsetsum} that there is a
  polynomial-time computable function $h$ such that $h(\psi)$ outputs
  an instance $X=\{x_1,\ldots, x_m\}$ and $T\in \mathbb{N}$ of
  \textsc{\#SubsetSum} such that
  \begin{align*}
   L & := \#\left\{ Y\subseteq X : \sum_{x\in Y} x = T \right\} \\
   & \phantom{:}= \#\left\{ (b_1,\ldots, b_n)\in \{0,1\}^n : \psi(b_1,\ldots,b_n)=1 \right\}.
  \end{align*}

  Using binary search, with at most $2\cdot m$ queries to the
  \textsc{$K$th largest subset} problem we can compute
  \begin{align*}
    M := \#\left\{ Y\subseteq X : \sum_{x\in Y} x \le T \right\}
    \quad \text{and}\quad
    N := \#\left\{ Y\subseteq X : \sum_{x\in Y} x \le T-1 \right\}.
  \end{align*}
  We now have that $L=M-N$, and hence a majority of the variable
  assignments satisfies $\psi$ if and only if $M-N \ge 2^{n-1}+1$.

  In the same fashion, we can show hardness of the \textsc{$K$th
    largest $m$-tuple} problem. Set $X_i := \{ 0, x_i\}$ for every
  $1\le i \le m$. As before, using binary search with at most $2\cdot
  m$ queries we can compute
  \begin{align*}
    M & := \#\left\{ (x_1,\ldots,x_m)\in X_1\times \cdots \times X_m : \sum_{i=1}^m
    x_i \ge T\right \}\\ & \phantom{:} = \#\left\{ Y\subseteq X : \sum_{x\in Y} x \ge T
    \right\}
  \end{align*}
  and
  \begin{align*}
    N & : = \#\left\{ (x_1,\ldots,x_m)\in X_1\times \cdots \times X_m : \sum_{i=1}^m
    x_i \ge T + 1 \right\}\\ & \phantom{:} =
    \#\left\{ Y\subseteq X : \sum_{x\in Y} x \ge T + 1 \right\}.
  \end{align*}
  As above, $L=M-N$ and a majority of the variable assignments
  satisfies $\psi$ if and only if $M-N \ge 2^{n-1}+1$, from which
  PP-hardness of the problem follows.\qed
\end{prf}


\section{Applications} \label{sec-other-problems}

In this section, we discuss several problems whose NP-hardness has
been shown in the literature by reductions from the \textsc{$K$th
  largest subset} (or \textsc{$m$-tuple}) problem.  By
Theorem~\ref{thm-main}, all of these problems become PP-hard. As
mentioned in Section~\ref{sec-intro}, Toda's Theorem implies that if
any of these problems were in NP (and hence NP-complete), the
polynomial-time hierarchy would collapse to P$^\text{NP}$.

\subsection{The \textsc{$K$th largest-area convex polygon} Problem}

Chazelle~\cite{Cha85} posed the following question: ``[G]iven $n$
points in the Euclidean plane, how hard is it to compute the $K$th
largest-area convex polygon formed by any subset of the points?''
This question was essentially answered by Salowe~\cite{Sal89}. In
particular, the following decision variant was considered
in~\cite{Sal89}:

\decision{$K$th largest-area convex polygon}
{A set $P$ of $n$ points $(x_1, x_2) \in \N \times \N$, and $K, B \in \N$.}
{Are there at least~$K$ subsets $P' \subseteq P$ for which $P' = \mathit{extr}( \mathit{conv} (P'))$ and the area of
$\mathit{conv} (P')$ is at least~$B$~?}
Here, $\mathit{conv}(P')$ refers to the convex hull of the points in~$P'$, and $\mathit{extr}( \mathit{conv} (P'))$ refers to the set of extreme points in that convex hull.
Salowe~\cite{Sal89} showed that the \textsc{$K$th largest-area convex polygon} problem is NP-hard, via a reduction from the \textsc{$K$th largest $m$-tuple} problem.
In fact, we have:
\begin{thm} \label{thm-convex-polygon}
The \textsc{$K$th largest-area convex polygon} problem is PP-complete under polynomial-time Turing reductions.
\end{thm}
\begin{prf}
PP-hardness follows from the reduction in~\cite{Sal89} combined with Theorem~\ref{thm-main}.
Towards membership in~PP, construct a PP-TM $M$, which, given an instance of the \textsc{$K$th largest-area convex polygon} problem, selects a subset~$P'$ of~$P$ uniformly at random, and then checks (in polynomial time) whether $P' = \mathit{extr}( \mathit{conv} (P'))$ and the area of
$\mathit{conv} (P')$ is at least~$B$.
If yes, $M$ accepts, otherwise $M$ rejects.
So the acceptance probability is at least $K/2^n$ if and only if
the given instance of the \textsc{$K$th largest-area convex polygon} problem is a yes-instance.
An application of Lemma~\ref{lem-PP-threshold} shows membership in~PP.
\qed
\end{prf}
Salowe~\cite{Sal89} remarked that the corresponding enumeration problem is \#P-complete.
From this observation it also follows that the \textsc{$K$th largest-area convex polygon} problem is PP-hard, although Salowe~\cite{Sal89} only claims NP-hardness.
He also says ``it is not clear that [the \textsc{$K$th largest-area convex polygon} problem] is in~NP''.
But this seems unlikely in view of Theorem~\ref{thm-convex-polygon}.

\subsection{The \textsc{vertex counting} Problem}

Let $m,n \in \N$ with $m > n$.  Let $b \in \Q^m$ be a column vector
and $A \in \Q^{m \times n}$ be a rational $m\times n$ matrix with its
rows denoted by $a_1, \ldots, a_m \in \Q^n$.  Then $A$ and $b$ define
a polyhedron $P := \{x \in \R^n : A \cdot x \le b\}$.
A \emph{vertex} of~$P$ is a point that is the unique intersection of
at least $n$ of the bounding hyperplanes $a_i \cdot x = b_i$, $i \in
\{1, \ldots, m\}$.  The following problem was considered by
Dyer~\cite{vertexEnumeration}:

\decision{vertex counting}
{A matrix $A\in \mathbb{Q}^{m\times n}$, $b\in \mathbb{Q}^n$, 
  and $K \in \N$.}
{Does the polyhedron $A\cdot x\le b$ have at most $K$ vertices~?}
Dyer~\cite[Proposition~3]{vertexEnumeration} has shown that the
\textsc{vertex counting} problem is NP-hard by a Turing reduction from
the \textsc{$K$th largest subset} problem. In fact, we have:

\begin{thm}
 The \textsc{vertex counting} problem is PP-complete under polynomial-time Turing reductions.
\end{thm}
\begin{prf}
PP-hardness follows from the reduction in~\cite{vertexEnumeration}
combined with Theorem~\ref{thm-main}.  To prove membership in PP, we
use the fact that PP is closed under complement~\cite{Gill77:PP}.  So
it suffices to prove PP-membership of the \textsc{co-vertex counting}
problem, which is like the \textsc{vertex counting} problem except
that it asks whether the polyhedron has \emph{at least} $K$ vertices.
Construct a PP-TM~$M$, which, given an instance of the
\textsc{co-vertex counting} problem, selects $n$ bounding hyperplanes,
say $h_1, \ldots, h_n$, uniformly at random, and then checks whether
they have a unique intersection.  If no, $M$ rejects.  Otherwise $M$
computes the unique intersection, say~$p$, and checks whether $p$ is
in the polyhedron.  If no, $M$ rejects.  Otherwise $M$ computes the
set~$S$ of all bounding hyperplanes on which $p$ lies.  Then $M$
computes the lexicographically smallest $n$-element subset~$S'$ of~$S$
such that $p$ is the unique intersection of the hyperplanes in~$S'$.
That set~$S'$ can be computed in polynomial time, e.g., by greedily
adding hyperplanes with linearly independent normal vectors.  If $S' =
\{h_1, \ldots, h_n\}$, then $M$ accepts, otherwise $M$ rejects.  By
construction, any vertex~$p$ is accepted by exactly one run of~$M$.
So the acceptance probability is at least $K \big/ {m \choose n}$ if
and only if the given instance of the \textsc{co-vertex counting}
problem is a yes-instance.  An application of
Lemma~\ref{lem-PP-threshold} shows membership in~PP.  \qed
\end{prf}

\subsection{Routing Flows with Delay Guarantees}

In~\cite{QoSRouting} the problem of routing flows through networks is considered, where the information available for making routing decisions is probabilistic.
The aim of the routing algorithm is to guarantee certain quality-of-service requirements.
One of the specific problems is to evaluate the \emph{end-to-end delay} of a given path, as we describe in the following.

Let $e_1, \ldots, e_n$ be a path in a network.
For each $i \in \{1, \ldots, n\}$, the number $d_i \in \N_0$ is a random \emph{delay} introduced by the link~$e_i$.
For all $i \in \{1, \ldots, n\}$ and $k \in \N_0$, the probability that $d_i = k$ is given by $p_{i,k} \in [0,1] \cap \Q$.
It is assumed that for each $i$ there are only finitely many $k$ with $p_{i,k} > 0$, and that the delays $d_i$ are independent.
The \emph{end-to-end delay} is $\sum_{i=1}^n d_i$.
An \emph{end-to-end delay requirement} is given by a maximum delay $D \in \N_0$ and a probability threshold $\tau \in [0,1] \cap \Q$.
Guerin and Orda refer to the following problem as ``problem $P(\pi)$''~\cite{QoSRouting}:

\decision{delay guarantee}
{A path length $n \in \N$, a list of all positive $p_{i,k} \in [0,1] \cap \Q$, the maximum delay $D \in \N_0$, a probability threshold~$\tau \in [0,1] \cap \Q$.}
{Is the probability of $\sum_{i=1}^n d_i \le D$ at least~$\tau$~?}
In~\cite[Lemma IV.1]{QoSRouting} the \textsc{delay guarantee} problem was shown NP-hard by a reduction from the \textsc{$K$th largest subset} problem.
In fact, we have:

\begin{thm}
 The \textsc{delay guarantee} problem is PP-complete under polynomial-time Turing reductions.
\end{thm}
\begin{prf}
PP-hardness follows from the reduction in~\cite{QoSRouting} combined with Theorem~\ref{thm-main}.
Towards membership in~PP, construct a PP-TM $M$, which, given an instance of the \textsc{delay guarantee} problem, takes independent random values for the~$d_i$ according to the given~$p_{i,k}$ and then checks whether $\sum_{i=1}^n d_i \le D$.
If yes, $M$ accepts, otherwise $M$ rejects.
So the acceptance probability is at least~$\tau$ if and only if the given instance of the \textsc{delay guarantee} problem is a yes-instance.
An application of Lemma~\ref{lem-PP-threshold} shows membership in~PP.
\qed
\end{prf}

\subsection{Further Probabilistic Problems}

The \textsc{$K$th largest subset} problem has been used by Laroussinie
and Sproston~\cite{LSdurational05} to show NP-hardness and
coNP-hardness of model-checking ``durational probabilistic systems''
(a variant of timed automata) against the probabilistic and timed
temporal logic~\textsc{PTCTL}.  Combining this reduction (which is
similar to the one for the \textsc{delay guarantee} problem) with
Theorem~\ref{thm-main} strengthens the NP-hardness and coNP-hardness
results of~\cite{LSdurational05} to PP-hardness.

Finally, the \textsc{$K$th largest $m$-tuple} problem was used
by Cire et al.~\cite{CCvH12} to show NP-hardness of another probabilistic problem,
namely determining whether a so-called ``chance-alldifferent
constraint'' has a solution.  This is a weighted extension of
stochastic constraint programming.  Cire et al.~\cite{CCvH12} remarked:
``Also, we are not aware if the problem of deciding whether there
exists a feasible solution to chance-alldifferent is in NP.''  But
this seems unlikely, as their reduction combined with
Theorem~\ref{thm-main} proves PP-hardness of this problem.





\bibliographystyle{elsarticle-num-names}
\bibliography{references}


\end{document}